\documentclass[prd,preprintnumbers,showpacs,amsmath,amssymb,twocolumn,nofootinbib]{revtex4}
\usepackage{latexsym,ifthen,graphics,color,hyperref,epsfig}

\newcommand{\ie}{{i.e.}}
\newcommand{\wrt}{with respect to}

\newcommand{\Kahler}{K\"{a}hler}

\newcommand{\be}{\begin{equation}}
\newcommand{\ee}{\end{equation}}

\newcommand{\eq}[1]{(\ref{eq:#1})}
\newcommand{\eqs}[2]{(\ref{eq:#1}) and~(\ref{eq:#2})}

\newcommand{\der}[2]{\frac{d #1}{d #2}}

\newcommand{\anti}[1]{\overline{#1}}
\newcommand{\suf}{\Phi}
\newcommand{\asuf}{\anti{\Phi}}

\newcommand{\flow}{\Lambda \partial_\Lambda}

\newcommand{\hepth}[1]{hep-th/#1}
\newcommand{\hepph}[1]{hep-ph/#1}
\newcommand{\arXiv}[2]{arXiv:{#1} [#2]}

\begin{document}

\preprint{DIAS-STP-08-08}

\title{Constraints on an Asymptotic Safety Scenario for the Wess-Zumino Model}

\author{Oliver J.~Rosten}
\email{orosten@stp.dias.ie}
\affiliation{Dublin Institute for Advanced Studies, 10 Burlington Road, Dublin 4, Ireland}

\begin{abstract}
	Using the nonrenormalization theorem and Pohlmeyer's theorem, it is proven
	that there cannot be an asymptotic safety scenario for the Wess-Zumino model
	unless there exists a non-trivial fixed point with (i) a negative anomalous
	dimension (ii) a relevant direction belonging to the \Kahler\ potential.
\end{abstract}

\pacs{11.10.Gh,11.10.Hi,11.30.Pb}

\maketitle

In this note, we will consider the existence of certain
renormalization group fixed points in theories of
a chiral superfield. Suppose that a non-trivial fixed point
exists and, moreover, that there is a renormalized trajectory~\cite{Wilson} 
emanating from it, such that the
low energy effective theory is well described by the Wess-Zumino model.
It will be proven that, for such an asymptotic safety scenario~\cite{Weinberg-AS} 
to occur,  the putative
fixed point must have both a negative anomalous dimension\footnote{
It is worth pointing out that in the vicinity of
a nonperturbative fixed point, we cannot rule out a negative anomalous dimension,
$\gamma$, by the usual
unitarity arguments. These relate the unitarity constraint $0\leq Z \leq 1$ to a non-negative $\gamma$
via a perturbative calculation; but there is no reason to believe such a calculation at a nonperturbative
fixed point (see~\cite{Etsoku} for an interesting discussion on negative anomalous dimensions).
}
and at least one relevant operator belonging to the \Kahler\
potential.
This generalizes earlier work~\cite{Zumino-Trivial} 
on zeros of the
$\beta$-function of the Wess-Zumino model in a way that
will be precisely spelt out below.

To formulate our argument, we introduce the Wilsonian effective action, $S_\Lambda$,
constructed by
integrating out degrees of freedom between the bare scale and a lower,
effective scale, $\Lambda$ (this implies that we have transferred to Euclidean space, so that momenta can be readily separated into large and small).%
\footnote{To explicitly compute $S_\Lambda$ would require that we write down an exact renormalization group~\cite{Wilson}
equation. However, the following arguments are sufficiently general that all we need to do is
suppose that this can be done.} 
The Wilsonian effective action, being infrared safe, does not suffer from
the holomorphic anomaly in the massless case. Therefore, the nonrenormalization theorem always
holds and the superpotential does not renormalize, even nonperturbatively~\cite{nrts}.

To conveniently uncover fixed point behaviour, we rescale
to dimensionless variables by dividing all quantities (coordinates and fields) by $\Lambda$ 
raised to the 
appropriate scaling dimension. In the case of the chiral superfield, $\suf$, (and its conjugate) we must take
account of the anomalous scaling according to
\be
	\suf \rightarrow \suf \sqrt{Z} \Lambda,
\label{eq:rescale}
\ee
where $Z$ is the field strength renormalization and
the anomalous dimension is defined by
\be
	\gamma(\Lambda) \equiv \Lambda \der{\ln Z}{\Lambda}.
\ee
As a consequence of the rescalings, the superpotential does now renormalize,
but just according to the (anomalous) mass dimension of the various couplings.
In particular,
denoting the rescaled three-point superpotential coupling by $\lambda(\Lambda)$, we have that
\be
	 \beta_\lambda \equiv \Lambda \der{\lambda}{\Lambda} 
	 =
	 \frac{3\lambda  \gamma}{2}.
\label{eq:beta}
\ee

In the rescaled variables, a fixed point is defined by
\be
	\flow S_\star[\asuf,\suf] = 0,
\label{eq:FP}
\ee
where $\flow$ is performed at constant $\asuf, \suf$ and a star is used to denote
a fixed point quantity (it is emphasised that a fixed-point action is something which is \emph{solved} for, using an exact renormalization group equation, not something which is chosen by hand). Immediately, it is apparent from~\eqs{beta}{FP}
that if $\lambda_\star \neq 0$,
 then it must be that $\gamma_\star = 0$. 
 
However, there is a theorem due to Pohlmeyer~\cite{Pohlmeyer} which
tells us that, if the two-point function in a scale invariant theory is canonical---\ie\ the anomalous  dimension is zero---then the field is a massless free field. Therefore, in the
current scenario, the only  fixed point (\ie\ scale invariant) theory with $\gamma_\star = 0$
must correspond to the Gaussian fixed point. 
This was the reasoning used in~\cite{Zumino-Trivial}
to rule out non-trivial zeros of the $\beta$-function in the Wess-Zumino model;
the same logic has also been applied to the O$(N)$ symmetric Wess-Zumino
model~\cite{Nappi}. Here, though, we deal with general fixed point actions.

However, the condition that $\lambda_\star = 0$ is
not sufficient to rule out an asymptotic safety scenario for
the Wess-Zumino model. This is because, although a putative non-trivial
fixed point cannot possess a three-point superpotential term, it could
be that (i) $\lambda$ constitutes a relevant direction at the fixed point
(ii) trajectories  initiated along the $\lambda$ direction happen to flow
towards the Gaussian fixed point. Note that a marginally relevant $\lambda$ will not do,
because this requires $\gamma_\star = 0$ and we again fall foul of Pohlmeyer's theorem.

Let us suppose that such a scenario
is realized: the non-trivial fixed point action is perturbed in the $\lambda$ direction, inducing
a flow
towards the Gaussian fixed point. Now, in the vicinity of the Gaussian fixed point
the low energy effective theory is described arbitrarily well by the Wess-Zumino
model. This follows simply because, although $\lambda$ is irrelevant \wrt\ the
Gaussian fixed point, it is only marginally so,  and so all other couplings
(besides the mass, which can
be ignored in this discussion) die off much faster.

Trajectories which emanate from fixed points are called renormalized trajectories~\cite{Wilson}.
As straightforwardly shown in~\cite{TRM-Elements}, a renormalized trajectory is such that all scale dependence of the action appears through (i) the relevant couplings with which the fixed point action has been perturbed (ii) the anomalous dimension of the field. This is referred to a `self-similarity'~\cite{Shirkov,TRM-Elements}; it is worth noting that self-similarity is a 
nonperturbative statement of renormalizability~\cite{TRM-Elements}.
In the current context, we have supposed that the fixed point action has been perturbed in the $\lambda$-direction. Were it not for the nonrenormalization theorem, we would expect the action along the resulting renormalized trajectory to depend on both $\lambda(\Lambda)$ and $\gamma(\Lambda)$. However, the two quantities are related by~\eq{beta} and so we can write simply
\be
	S_\Lambda[\asuf,\suf] = S[\asuf,\suf](\gamma(\Lambda)).
\label{eq:self-similar}
\ee

As just stated, in order to construct this renormalized trajectory, it must be that $\lambda(\Lambda)$
is relevant
\wrt\ the non-trivial fixed point. This requires that $\gamma_\star <0$, as follows from~\eq{beta}.
Crucially, however, sufficiently close to the Gaussian fixed point---where we
can rely on perturbation theory done with the Wess-Zumino model---we know that
the anomalous dimension is positive.

Therefore, in going from the UV fixed point down to
the vicinity of the Gaussian fixed point,  $\gamma(\Lambda)$ must pass through zero 
(at least once). 
Consider the first time that this happens.
Since all scale dependence along our renormalized trajectory is carried by $\gamma(\Lambda)$ then,  if 
$\gamma(\Lambda)$ ever vanishes, we must be at a fixed point.
Now, on the one hand, this fixed point cannot be the Gaussian one: the 
action in the vicinity of the Gaussian fixed point is (essentially) the Wess-Zumino action, 
but  $\gamma(\Lambda)$ has not yet increased above zero, by assumption. 
On the other hand, Pohlmeyer's theorem tells
us that this fixed point cannot be anything else! Therefore, our original assumption that
there exists a non-trivial fixed point with a trajectory, spawned along the
$\lambda$ direction,  emanating from it such that the
low energy effective theory is well described by the Wess-Zumino model,
must be incorrect. 

However, suppose that the fixed point also possesses a relevant operator
coming from the \Kahler\ potential, $\mathcal{O}[\asuf,\suf]$, with coupling
$g(\Lambda)$ (obviously, we can generalize this to several such operators). 
Perturbing the fixed point action in both the $\lambda$ and $g$ directions,
the action along the resulting renormalized trajectory now reads
\be
	S_\Lambda[\asuf,\suf] =  S[\asuf,\suf](g(\Lambda),\gamma(\Lambda)).
\ee
Whilst it is still true that, in order for an asymptotic safety scenario
to be realized for the Wess-Zumino model, the anomalous dimension
must pass through zero, it is no longer true that the vanishing of
$\gamma(\Lambda)$ at some scale 
necessarily corresponds to fixed point, since $g(\Lambda)$ could
still be flowing.

Assuming such an asymptotic safety scenario to exist, we now have
the following picture of the renormalization group flows. If we perturb away from the
non-trivial fixed point in just the $\lambda$ direction, then we must
shoot off away from the Gaussian fixed point. (A finite distance along
the resulting
trajectory, it may be that  $\mathcal{O}[\asuf,\suf]$ is generated, but
now we have $g(\Lambda) = g(\gamma(\Lambda))$.) However, by
appropriately perturbing the fixed point in both the $\lambda$ and $g$
directions, we flow towards the Gaussian fixed point, with the low
energy effective action being well described by the Wess-Zumino
action.
The question as to whether such non-trivial fixed points actually exist
will be addressed in a companion paper~\cite{WIP}.

\vspace{-0.4ex}

\begin{acknowledgments}
I acknowledge IRCSET for financial support. I would like to thank Denjoe O'Connor for
comments on the manuscript and, particularly, Werner Nahm for a very useful
discussion.
\end{acknowledgments}

\providecommand{\href}[2]{#2}\begingroup\raggedright\endgroup

\end{document}